\documentclass{svmult} 
\usepackage{makeidx}         
\usepackage{graphicx}        
\usepackage{multicol}        
\usepackage[bottom]{footmisc}
\makeindex             

\newcommand{\invitro}{\textit{in vitro} }
\newcommand{\ie}{i.e.\ }
\newcommand{\mycite}[1]{\cite{#1}}\newcommand{\fig}[1]{Fig.~\ref{#1}}

\newcommand{\SSS}[1]{\scriptscriptstyle{#1}}
\newcommand{\mum}{\rm \mu m }
\newcommand{\nanom}{\rm nm }
\newcommand{\Np}{N_+}\newcommand{\Nm}{N_-}\newcommand{\np}{n_+}\newcommand{\nm}{n_-}
\newcommand{\Fs}{F_s}\newcommand{\Fd}{F_d}\newcommand{\epsO}{\epsilon_0}\newcommand{\piO}{\pi_0}
\newcommand{\vF}{v_{\SSS \rm F}}\newcommand{\vB}{v_{\SSS\rm B}}

\newcommand{\FP}{F_{+}}\newcommand{\FM}{F_{-}}

\renewcommand{\P}{(+)}\newcommand{\M}{(-)}\newcommand{\N}{(0)}
\newcommand{\PM}{(-+)}\newcommand{\PMN}{(-0+)}

\begin{document}

\title*{Traffic by small teams of molecular motors}
\author{Melanie J.I. M{\"u}ller\inst{1} \and Janina Beeg\inst{1} \and Rumiana Dimova\inst{1} \and Stefan Klumpp\inst{2} \and Reinhard Lipowsky\inst{1}}
\institute{Max Planck Institute of Colloids and Interfaces, Science Park Golm, 14424 Potsdam, Germany.
\texttt{mmueller@mpikg.mpg.de}
\and Center for Theoretical Biological Physics, University of California San Diego, La Jolla, CA 92093-0374, USA.}
\maketitle


\textbf{Abstract}. Molecular motors transport various cargos along cytoskeletal filaments, analogous to trucks on roads. In contrast to vehicles, however, molecular motors do not work alone but in small teams. We describe a simple model for the transport of a cargo by one team of motors and by two teams of motors, which walk into opposite directions. The cooperation of one team of motors generates long-range transport, which we observed experimentally \invitro. Transport by two teams of motors leads to a variety of bidirectional motility behaviour and to dynamic instabilities reminiscent of spontaneous symmetry breaking. We also discuss how cargo transport by teams of motors allows the cell to generate robust long-range bidirectional transport. 


\section{Introduction: Traffic of molecular motors and the need for motor teams}

Molecular motors are protein molecules which power various transport processes in cells \mycite{Howard01}. Their traffic is in many ways similar to car traffic. While cars drive on roads, molecular motors walk along tracks provided by cytoskeletal filaments. While cars consume petrol, molecular motors use energy from the hydrolysis of adenosine triphosphate (ATP) in order to perform mechanical work. Prominent examples are the kinesin and dynein motors traveling along filaments called microtubules which form a 'highway network' inside the cell. These are 'one-way highways': dynein motors walk preferentially to one end of the microtubules (called 'minus' end) while most kinesins walk to the opposite microtubule 'plus' end.  Just like cars, motors \textit{can} walk backwards, but are rather bad at it: they usually do so only slowly and if forced. Of course, there are also important differences between cars and motors. An obvious difference is the length scale: While cars are  several m in size and travel km distances, molecular motors are only about $100\,\nanom$ in size and travel $\mum$ distances. As a consequence, molecular motors work in an environment dominated by thermal noise, which for a car would be comparable to permanently driving in a hurricane. This leads to another interesting feature of molecular motors: they can 'fly', \ie unbind from their track. However, upon unbinding they lose their ability to perform directed motion and randomly diffuse in the surrounding solution until they finally rebind to a filament. Due to these unique features, the traffic of molecular motors has become an attractive problem for traffic modeling and studies of non-equilibrium transport \mycite{LipowskyKlumpp05LifeIsMotion,LipowskyNieuwenhuizen01PRLRandomWalksCompartments,ParmeggianiFrey03}.

In this paper we address another aspect of cargo transport by molecular motors. While in road traffic a single truck usually suffices to carry its cargo, in cellular traffic this is not so. As mentioned, due to thermal noise molecular motors unbind from their track from time to time. For the molecular motor kinesin this happens on average after a 'run length' of about $1\,\mum$. A cellular cargo, however, must accomplish distances of tens of $\mum$, and in some extremely large cells like neurons even up to a metre \mycite{GoldsteinYang00}.  Furthermore, the cellular surrounding is very viscous, leading to high frictional forces which can become too large for a single motor. A third problem for the cell is bidirectional transport. In a cell, the 'one-way' microtubule tracks are usually arranged in an isotropic way, pointing from the cell centre to the cell periphery \mycite{LaneAllan98}. A single motor walks in only one direction along these tracks. However, many cellular cargos travel bidirectionally \mycite{Gross04,Welte04}, as has to be the case in order not to accumulate cargo at either the cell periphery or the cell centre \mycite{KlumppLipowsky05FilamentArrays,MullerLipowsky05HalfOpenTube}.

The cell solves all three problems by using several molecular motors rather than a single motor to transport a cargo. In order to obtain large run lengths and forces, several motors of the same species can work together as 'one team'. In order to accomplish bidirectional transport, motors with different directionalities transport a single cargo as 'two teams'. The number of motors  in a team is small, typically between 1 and 10 \mycite{HabermannBurkhardt01,GrossShubeita07}. In this paper we review recent theoretical analyses \mycite{KlumppLipowsky05PNASCargoTransport,MullerLipowsky07TugOfWar} and \invitro experiments \mycite{BeegLipowsky07} performed in our group that studied the cooperation of small teams of molecular motors pulling a single cargo along a unidirectional filament network. After defining our model, we will first examine the transport by one cooperating team of molecular motors of the same species and then the transport by two antagonistic teams of molecular motors that walk into opposite directions.

\section{Theoretical modelling: from one to many motors}

We consider a cargo which is transported by fixed numbers of $\Np$ plus and $\Nm$ minus motors. Because of thermal fluctuations, a motor stochastically unbinds after some time. Therefore the cargo is pulled by a fluctuating number of motors, see \fig{fig:FluctuatingMotorsCombi}(b). The state of the cargo is determined by the numbers $\np$ and $\nm$ of pulling plus and minus motors. In the simplest case, the motors work independently of each other, and one can deduce the cargo behaviour from the behaviour of a single motor. A single motor can bind to the filament with the binding rate $\piO$, walk along it with the forward velocity $\vF$, and unbind from it with rate $\epsO$. If the motor has to work under a force $F$, which can be caused  by opposing motors, Stokes friction, or an optical trap, these rates become force-dependent. The unbinding rate increases exponentially with the forces \mycite{SchnitzerBlock00} as $\epsilon(F)=\epsO\,\exp(F/\Fd)$, where the force scale is set by the detachment force $\Fd$. The velocity decreases linearly \mycite{CarterCross05}, $v(F)=\vF(1-F/\Fs)$ until it reaches zero at the stall force $\Fs$. For higher loads, the motor walks backwards \mycite{CarterCross05} with the velocity $v(F)=-\vB(1-F/\Fs)$, with a very small backward velocity $\vB$.  

\section{Transport by one team of motors}

\begin{figure}[tb]\centering
\includegraphics[width=\textwidth]{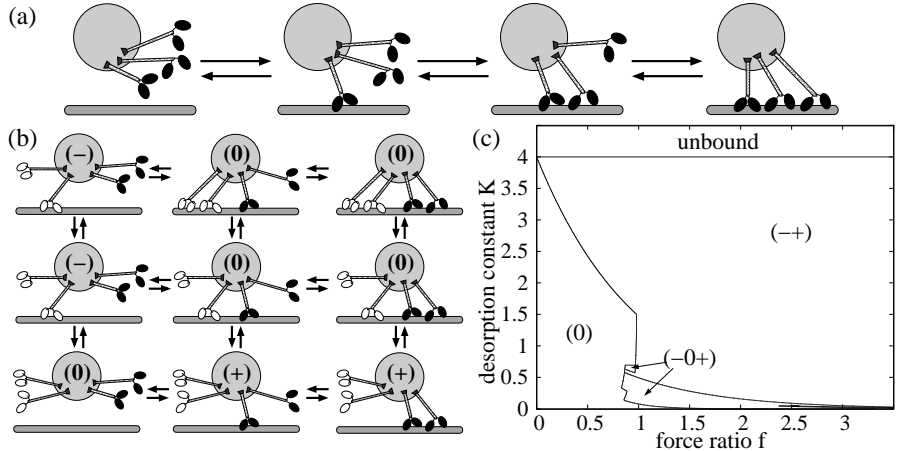}
\caption{Transport by small teams of motors.
(a) A cargo with $N=3$ motors is pulled by a fluctuating number of motors.
(b) A cargo with $\Np=2$ (black) plus motors and $\Nm=2$ (white) minus motors is pulled by a fluctuating number of plus and minus motors. States with only plus motors bound $\P$, only minus motors bound $\M$ and both types of motors bound $\N$ correspond to fast plus, fast minus and slow motion, respectively.
(c) Motility diagram for the symmetric tug-of-war of 4 against 4 motors with the same single-motor parameters (except their preferred direction). Depending on the single-motor force ratio $f=\Fs/\Fd$ of stall and detachment force and desorption constant $K=\epsO/\piO$, the cargo is in one of three motility states $\N$, $\PM$ or $\PMN$ as explained in the text. For high desorption constants, the cargo is unbound.}%
\label{fig:FluctuatingMotorsCombi}
\end{figure}

We first consider a cargo transported by $N=\Np$ plus motors and no minus motors, $\Nm=0$. The number $n=\np$ of motors which are bound to the filament fluctuates between $0$ and $N$, see \fig{fig:FluctuatingMotorsCombi}(a). As the motors work independently, the unbinding and binding rates for one motor in the cargo state with $n$ pulling motors are simply $n\epsO$ and $(N-n)\piO$ with the single motor unbinding rate $\epsO$ and binding rate $\piO$. This leads to a Markov process on the states $n=1,\ldots,N$, for which we have obtained a number of analytical results \mycite{KlumppLipowsky05PNASCargoTransport}.

In particular, the average run length, \ie the distance a cargo moves along a filament before it unbinds from it, increases essentially exponentially with the motor number $N$. This is because the cargo particle continues to move along the filament unless all $N$ motors unbind simultaneously. When the cargo is transported by the molecular motor kinesin, 3 kinesins suffice to cross a cell of $50\,\mum$ diameter, and 7-8 kinesins lead to average run lengths in the centimetre range \mycite{KlumppLipowsky05PNASCargoTransport}. The corresponding run length probability distribution is a sum of exponentials, which develops fat tails for large motor numbers $N$.

The increase of run length with increasing motor number has been observed \invitro \mycite{BlockSchnapp90,CoyHoward99,SeitzSurrey06}, but it has been difficult to determine the number of motors pulling the cargo. To overcome this limitation, we have recently used a combination of dynamic light scattering (DLS) and a comparison of measured and theoretical run length distributions to determine the number of pulling motors \mycite{BeegLipowsky07}. In our experiments, latex beads were incubated in solutions with different concentrations of kinesin motors. The kinesins bound stochastically to the beads and pulled them along an array of immobilized isopolar microtubules within a glass channel \mycite{BohmUnger01}. In such an assay, the maximal number of motors which are available for binding to the microtubule and pulling the cargo is not constant, but varies from bead to bead. The theoretical run lengths were therefore weighted with a truncated Poissonian distribution \mycite{BeegLipowsky07}, and then fitted to the measured run length distributions for 9 different kinesin concentrations using only 2 fit parameters. The agreement of theory and experiment allowed to calculate the maximal number $N$ of motors which were available for bead transport. This result was found to correspond well to the motor number independently estimated from the DLS measurement. 

\section{Transport by two teams of motors}

Next we consider a cargo with $\Np$ plus and $\Nm$ minus motors attached. The numbers $\np$ and $\nm$ of bound plus and minus motors, change stochastically as shown in \fig{fig:FluctuatingMotorsCombi}(b). As the motors are assumed to act independently, the rates for unbinding and binding of a single motor when the cargo is in the state $(\np,\nm)$ can be deduced from the corresponding single motor rates. The opposing motors exert force on each other, so that each plus motor feels the force $\FP$ and each minus motor the force $\FM$. Newton's third law requires $\np\FP=\nm\FM$. Furthermore, as both motor types are bound to the same cargo, the velocity of each plus motor under force $\FP$ must equal the velocity of each minus motor under force $\FM$. The force and velocity balance determine the motor forces $\FP$ and $\FM$ and therefore the motor binding and unbinding rates and the cargo velocity \mycite{MullerLipowsky07TugOfWar}.

It is instructive to consider the symmetric tug-of-war of $\Np=\Nm$  plus and minus motors which have the same single motor parameters and differ only in their preferred direction. In this case the direction of motion in each of the cargo states shown in \fig{fig:FluctuatingMotorsCombi}(b) is simply given by the majority motor type and is indicated by (+), (-) or (0) for plus motion, minus motion and slow motion, respectively. The probability $p(\np,\nm)$ that the cargo is in the state with $\np$ bound plus and $\nm$ bound minus motors can have either 1, 2 or 3 maxima, depending on the single motor parameters, see  \fig{fig:FluctuatingMotorsCombi}(c). As the cargo spends most of its time in configurations with high probability, these maxima characterize the large-scale cargo motion. For 'weak' motors with a low ratio $f=\Fs/\Fd$ of stall force to detachment force, the probability distribution  $p(\np,\nm)$ has only one maximum at a configuration with $\np=\nm$, which corresponds to no motion [no motion motility state $\N$]. When the motors have a high force ratio $f$, on the contrary, the motor number probabilities exhibit two maxima at $(n,0)$ and $(0,n)$. In this parameter range, the cargo switches stochastically between fast plus and fast minus motion [$\PM$ motility state]. In an intermediate range of $f$, all three types of maxima appear, and the cargo switches between fast plus motion, fast minus motion and pauses [$\PMN$ motility state].

For large force ratios $f$, the appearance of two maxima at $(n,0)$ and $(0,n)$ in a situation symmetric with respect to plus and minus motors is reminiscent of spontaneous symmetry breaking during continuous phase transitions. The reason for its appearance is a dynamic instability caused by the nonlinearity in the force-dependence of the single motor unbinding rate. The time for switching between the two non-symmetric maxima increases exponentially with the motor number $\Np=\Nm$, indicating a non-equilibrium phase transition in the infinite system.

If the tug-of-war is non-symmetric, the dynamic instability persists, and the cargo switches stochastically between fast plus motion, minus motion and / or pauses. However, now the plus-minus motor symmetry is lost, biased plus or minus motion is possible. Thus, even though the motors are engaged in a tug-of-war, fast motion into plus or minus direction can be generated.

\section{Discussion: robustness and regulation}

Why should cells use a team of motors instead of one strong motor which rarely unbinds from the filament? And why should cells use two teams engaged in a tug-of-war instead of one team only which is substituted by a team of opposite-directional motors when appropriate? The reasons may be robustness and sensitivity to regulation. A team of motors is more robust against failure of a single motor. And a team can be easily regulated by simply regulating the number of motors involved in the team. Two teams of motors can carry a cargo into two directions instead of only one. A bidirectionally moving cargo can search for its target, bypass obstacles and correct targeting errors. Furthermore, a cargo with two teams of motors engaged in a tug-of-war is very sensitive to regulatory mechanisms: because of the dynamic instability, small changes in the molecular properties (or of the number) of one or both motor types can qualitatively change the characteristics of cargo motion. The cargo can move into one direction faster or for a longer time and it may show net plus or net minus motion. In this way the cell can easily target its cargos as appropriate.

In summary, we have described a simple model for cargo transport by one or two teams of molecular motors. Despite its simplicity, the model exhibits a rich variety of motility behaviours and explains how the cell might satisfy its need for long-range bidirectional transport. 

\textbf{Acknowledgements}. The authors thank R. Serral Graci{\`a} and E. Unger for enjoyable collaborations. SK was supported by Deutsche Forschungsgemeinschaft (Grants KL818/1-1 and 1-2) and by the National Science Foundation through the Physics Frontiers Center-sponsored Center for Theoretical Biological Physics (Grants PHY-0216576 and PHY-0225630).



\printindex
\end{document}